\documentclass[preprint,superscriptaddress,amsmath,amsthm,amssymb,aps,pre,floatfix]{revtex4-1}

\usepackage[T1]{fontenc}
\usepackage{amssymb,amsmath}
\usepackage{amsfonts}
\usepackage{graphicx}
\usepackage{xcolor}
\usepackage{mathbbol}
\usepackage{bbm}

\usepackage{dcolumn}
\usepackage{hyperref}

\usepackage{bm}
\usepackage{dsfont}

\def\ket#1{|#1\rangle}

 \newcommand{\bs}{{\bf s}}
\newcommand{\bq}{{\bf q}}
\newcommand{\bp}{{\bf p}}

\newcommand{\bbx}{{\bf x}}
\newcommand{\bxi}{{\bm \xi}}

\newcommand{\gt}{g_t}
\newcommand{\LLc}{{\mathcal{L}}_c}

\newcommand{\abscisse}{u}

\newcommand{\dK}[1]{d^K \! {#1}}

\newcommand{\Sq}{S}
\newcommand{\Sx}{\mathcal{S}}

\newcommand{\LL}{\mathcal{L}}

\newcommand{\Vol}{\mathcal{V}}

\def\calm{\mathcal{M}}
\def\cA{\,\mathfrak{A}}
\def\cala{\mathcal{A}}

\def\hatO{\hat{O}}

\def\idmat{\mathds{1}}

\def\q0{\bq_\textrm{i}}
\def\p0{\bp_\textrm{i}}

\def\xpar{x_{\parallel}}
\def\xper{x_{\perp}}
\def\bxpar{{\bf x_{\parallel}}}
\def\bxper{{\bf        x_{\perp}}}
\def\bxsheer{{{\bf \hat x_{\perp}}}}
\def\xipar{\xi_{\parallel}}

\def\Ai{\textrm{Ai}}

\def\x0{{\bf x}_\textrm{i}}

\def\bx0{{\bf x}_\textrm{0}}

\newcommand{\rem}[1]{}
\newcommand{\chckout}[1]{}

\begin{document}

\begin{center}{\Large \textbf{Semiclassical evaluation of expectation values}
}\end{center}

\begin{center}
K.~M.~Mittal\textsuperscript{1,2},
O.~Giraud\textsuperscript{1},
D.~Ullmo\textsuperscript{1*}
\end{center}

\begin{center}
{\bf 1} Universit\'e Paris-Saclay, CNRS, LPTMS, 91405, Orsay, France
\\
{\bf 2} Indian Institute of Science Education and Research, Dr.~Homi Bhabha Road, Pune 411 008, India
\\
* denis.ullmo@universite-paris-saclay.fr
\end{center}

\begin{center}
\today
\end{center}

\section*{Abstract}
{\bf
Semiclassical mechanics allows for a description of quantum systems
which preserves their phase information, and thus interference
effects, while using only the system's classical dynamics as an
input.  In particular one of the strengths of a semiclassical
description is to present a coherent picture which (up to negligible
higher-order $\hbar$ corrections) is independent of the particular
canonical coordinates used.  This coherence relies however heavily on the
use of the stationary phase approximation, which appears in many
circumstances as the building block of the semiclassical approach.

It turns out however that in some (important) cases, a brutal
application of stationary phase approximation washes out all
interference (and thus quantum) effects.  This rather basic, but in the
end quite subtle, point has actually generated quite a bit of confusion
in the semiclassical literature.  In this paper, we address this issue
in detail in one of its simplest instantiations, namely the evaluation
of the time evolution of the expectation value of an operator. We
explain why it is necessary to include contributions which are not in
the neighborhood of stationary points and provide new semiclassical
expressions for the evolution of the expectation values.  Along the
way, we get a deeper understanding of the origin of these interference
effects and an intuitive geometric picture associated with them.

}

\section{Introduction}

Semiclassical approximations form a family of techniques, including
for instance the van Vleck approximation of the quantum
evolution operator \cite{VanVleck28}, 
the time-dependent WKB approximation for the propagation
of wave functions \cite{Dir47}, or the Gutzwiller trace formula for the
density of states \cite{Gut71, Gut89}, making it possible to describe
quantitatively a quantum system in a certain range of parameters,
using only an input from the corresponding classical dynamics.
In our nomenclature classical approximations of a quantum
quantity are approximations in which interference effects are ignored,
or lost. By contrast, {\em semiclassical} approximations
are such that
information about phases is kept, and thus provide a description of
interference effects.

The stationary phase approximation plays a central role in
semiclassical approaches.  Indeed, common tools in quantum mechanics,
such as the Fourier transform or the Wigner transform, involve
integrals of the form
\begin{equation} \label{eq:Sc-form}
\int \dK{\bq} \,A(\bq) e^{\frac{i}{\hbar} S(\bq)} \; ,
\end{equation}
where the integrand has a smooth envelope $A(\bq)$ and a rapidly
oscillating phase $S(\bq)/\hbar$.  For relatively deep reasons,
semiclassical approximations usually keep the leading-order term and
the first-order correction in $\hbar$, but neglect higher-order
corrections, which is exactly what the stationary phase approximation
does.  Within the semiclassical framework, integrals such as
\eqref{eq:Sc-form} are thus essentially always performed within the
stationary phase approximation, and the understanding that any such
integral has to be performed within this approximation actually
provides a global coherence to the semiclassical pictures. Consider
for example the van Vleck formula expressing the Feynman propagator as
a sum over classical trajectories. It can be derived through a WKB
approximation ``\`a la Maslov'' \cite{Maslov81}, where one looks for solutions of
partial differential equations in the form of formal series in
$\hbar$. It can also be obtained from the path integral representation
of the exact quantum evolution operator if all integrals in the path
integral formalism are performed within the stationary phase
approximation \cite{Schulman81}. Both approaches lead to exactly the same result.

One of the great strengths of semiclassical approaches is the
possibility to freely choose the system of coordinates suited to our
needs. Indeed, one can show that semiclassical approximations keep the
same form under a canonical change of the classical variables within
the stationary phase approximation.  This property is used in
particular by Maslov \cite{Maslov81} to treat cases where the amplitude $A(\bq)$ in
\eqref{eq:Sc-form} may become singular, as we shall see later in this
paper. 

Because of the identity between expressions obtained from the Maslov
$\hbar$ expansion and the stationary phase approximation, there is, in
the mind of many practitioners of the field, a kind of identification
between semiclassical and stationary phase approximations.  There are
of course some limitations to this identification, as in some
circumstances, e.g.~near turning points or caustics, or near the
bifurcation of a periodic orbit, a naive application of the stationary
phase approximation leads to divergences. These have to be cured by
uniform approximations, in which the behavior of the neighborhood of
stationary phase point is described more carefully than the usual
quadratic approximation.  It remains nevertheless that in almost all
circumstances, {\em only the neighborhood of the stationary phase
  points can contribute significantly} to an integral of the form
\eqref{eq:Sc-form}, and within a semiclassical approach to a physical
problem this is most of the time implicitly or explicitly assumed.

There is however a class of problems for which this rather natural
assumption turns out to fail.  This class includes relatively
``involved'' physical quantities, like fidelity \cite{JalabertPastawskiPRL2001,TomsovicCerrutiPRL2002,Tomsovic2016}, OTOC \cite{Klaus_OTOC_PRL2018,Jalabert_OTOC-PRE2018}, or the quantum
corrections to the conductance in mesocopic quantum dots
\cite{Klaus_Transport2002_PRL2002,Brouwer_transport_PRB2006}, but also
some others as benign as the time evolution of the expectation value
$\langle \hat O \rangle = \langle\Psi|\hat O|\Psi\rangle$ of an
operator $\hat O$ for some time-dependent quantum state
$\ket{\Psi_t}$ \cite{Ozo13}, or more general correlation functions
\cite{OzoBro16}.   In this paper, we shall focus on the simple case
of the expectation value.

To be more specific, consider a $K$-dimensional dynamical system, whose phase-space
$\calm$ is the set of points $\bbx=(\bp,\bq)$ with
$\bp=(p_1,\ldots,p_K)$ and $\bq=(q_1,\ldots,q_K)$. Suppose the system
is initially described by a state $\ket{\Psi_0}$, which evolves under
some Hamiltonian dynamics to a state $\ket{\Psi_t}$ at time $t$.  The
Wigner transform $O(\bbx)$ of an arbitrary operator $\hat O$ is
defined by
\begin{equation} 
\label{eq:Wigner}
O(\bbx)=\frac{1}{(2\pi\hbar)^K}\int \dK{\bq'} e^{-\frac{i}{\hbar}\bp.\bq'}\Big\langle
\bq+\frac{\bq'}{2}\Big|\hat O\Big|\bq-\frac{\bq'}{2}\Big\rangle, \qquad \bbx\in \calm.
\end{equation}
We denote by $W_t(\bbx)$ the Wigner transform of the state $\ket{\Psi_t}$, namely
\begin{equation} 
\label{eq:Wignerpsi}
W_t(\bbx)=\frac{1}{(2\pi\hbar)^K}\int \dK{\bq'} e^{-\frac{i}{\hbar}\bp.\bq'}\overline{\Psi_t}\left(\bq-\frac{\bq'}{2}\right)\Psi_t\left(\bq+\frac{\bq'}{2}\right)\,,
\end{equation}
where bar denotes complex conjugation.
The expectation value  $\langle \hat O \rangle(t) = \langle\Psi_t|\hat O|\Psi_t\rangle$ can then be written \cite{Ozo98} as the phase-space integral 
\begin{equation} 
\label{hata}
\langle \hat O \rangle(t) = \int \dK{\bbx} \, W_t(\bbx)O(\bbx).
\end{equation}
Suppose $\hat O$ is ``smooth'', in the sense that its Wigner transform varies only on the classical scale. Then one can show that keeping only the contribution of the neighborhood of stationary phase points in the integral of Eq.~\eqref{hata} leads to \cite{OzoRio02}
\begin{equation} \label{eq:TWA}
\langle \hat O \rangle_{\rm cl}(t) = \int \dK{\bbx}\, W_0(\bbx)\, O(\gt\bbx) = \int \dK{\bbx}\, W_0(\gt^{-1}\bbx)\, O(\bbx),
\end{equation}
with $\gt$ the classical Hamiltonian flow characterizing the classical
dynamics associated with the Hamiltonian $\hat H$. In words,
$\langle \hat O \rangle_{\rm cl}(t)$ is obtained as the overlap of the
Wigner transform of the operator $\hat O$ with the classically evolved
Wigner transform of the state $\ket{\Psi}$. This expression is often
referred to as the Linearized Semiclassical Initial Value
Representation (LSC-IVR) \cite{Miller98a} in chemical or molecular
physics, or the Truncated Wigner Approximation (TWA) in the context of
cold atoms \cite{Pol10}, but can be tracked back at least to the work
of Heller \cite{Hel76} in 1976 and is discussed in the review written
by Wigner and co-authors \cite{Wigner84} in 1984. Within this
approximation, all interference effects are washed out, and the
expression \eqref{eq:TWA} yields essentially the classical
approximation to the mean value.

This result actually leads us to consider several
possibilities. The first one would be that either interference
  effects are indeed completely washed out, and that for the
  expectation value {\em of any smooth operator}, the classical
  approximation Eq.~\eqref{eq:TWA} is indeed a correct description of
  the quantum evolution; or that rapidly oscillating corrections to
  this classical terms do exist but that they are beyond the reach of
  semiclassics.  There is however a large literature, starting from
  the seminal paper of Tomsovic and Heller
  \cite{TomsovicHeller_PRL1991}, and including some situations for which the issue of non-applicability of the stationary phase approximation exist 
  \cite{JalabertPastawskiPRL2001,TomsovicCerrutiPRL2002,Tomsovic2016,Klaus_Transport2002_PRL2002,Brouwer_transport_PRB2006,Klaus_decay_PRL2008,Klaus_OTOC_PRL2018,Jalabert_OTOC-PRE2018,Ozo19}, which shows that on
  a quite general basis the interference effects that set on after the
  Ehrenfest time can be described within a semiclassical approach, and
  that their effects is non negligible.  We therefore clearly also
  expect this to be the case for the mean value of smooth operators \cite{Ozo13,OzoBro16}.

  All the examples above provide a clear demonstration of the
    effectiveness of semiclassics.  However, the physical problem they
    describe, and the approach they use, often involve either a
    statistical argument \cite{JalabertPastawskiPRL2001,TomsovicCerrutiPRL2002,Tomsovic2016,Klaus_Transport2002_PRL2002,
      Brouwer_transport_PRB2006, Klaus_decay_PRL2008,
      Klaus_OTOC_PRL2018, Jalabert_OTOC-PRE2018} or, in the spirit of
    the Initial Value Representation (IVR), to perform numerically the
    integral for which the stationary phase approximation fails
    \cite{Ozo19}. Therefore, the rationale dictating when stationary
  phase can be used and when it cannot is not always absolutely clear,
  and one may be worried about the fact that this necessity to do
  without stationary phase approximation in some circumstances but not
  in others could harm the general coherence of the semiclassical
  approach. The goal of our work is to clarify these issues on the
  simple case of the operator mean value. Our aim is to provide a
  semiclassical description of interference effects for the time
  evolution \eqref{hata} of the mean value of a smooth operator
  $\hat O$. We shall assume that $\hat O$ is smooth in the sense that
  its Wigner transform varies only on classical scales, that is, shows
  no short-scale quantum features. In particular we will clarify  why
  in the integral of Eq.~\eqref{hata} one should keep contributions
  which are not in the neighborhood of a stationary phase point.
  Along the way, we shall gain a deeper understanding of the origin of
  the interference terms in the time-dependent expectation values of
  operators, as well as a new semiclassical expression for them which,
  for some (simple enough) operators $\hat O$, are rather explicit.

  The structure of the paper is as follows. We start, in the two
  following sections, by reviewing two important tools of
  semiclassical theory that we will need for our discussion. We begin
  in section~\ref{sec:SemiclassicWF} by a presentation of the Maslov
  approach \cite{Maslov81} which relates the semiclassical evolution
  of a quantum wavefunction to the classical evolution of the
  Lagrangien manifold on which it is constructed.  We then derive in
  section~\ref{scwf} an expression for the semiclassical Wigner
  function, which is the starting point of our discussion of the
  expectation value of smooth operators. This derivation will follow
  very closely the spirit of the original work of Berry
  \cite{Ber77}. However, we shall express this semiclassical Wigner
  function in a slightly different form, which transforms under
  canonical change of variable in a rather transparent way.  This will
  in particular make it possible to show that the relevant
  interference terms in the computation of a smooth operator's mean
  value are expressed only in terms of {\em canonical invariants.}
  Relying on this canonical invariance of the formalism, we then
  address in section~\ref{sec:expectation} the question of the
  interference terms in the expectation value of a smooth operator,
  discuss their origin, the precise place where stationary phase
  approximation cannot be used (and why), and provide an expression
  for these interference terms which turns out to be in the end rather
  intuitive and natural.  Finally, we conclude in section~\ref{ccl}
  with a comparison between the point of view taken in this paper and
  the approach suggested by an Initial Value Representation (IVR)
  approach.

\section{Semiclassical wavefunctions}
\label{sec:SemiclassicWF}

\subsection{Wavefunctions from Lagrangian manifolds}

We assume in this paper that the initial state we consider is a
semiclassical wavefunction, that is, a state of the form
$\psi_0(\bq)=\cala(\bq) \exp(i \Sq (\bq)/\hbar)$, or possibly a finite
sum of terms of that form. Such a function is the product of a smooth
envelope $\cala(\bq)$ and a function that oscillates rapidly when
$\hbar\ll 1$.  Provided one extend this
  notion to include complex Lagrangian manifolds so that coherent
  states can be treated within this framework \cite{Tomsovic2018}, all the quantum states
  of interest in semiclassical physics in practice fall in that category.

A procedure introduced by Maslov \cite{Maslov81}, that
we briefly sketch now, allows to associate such a function with a
Lagrangian manifold in the classical phase space. We stress that we
do not introduce a specific dynamics/Hamiltonian yet, and
the semiclassical wavefunction that we are going to build from the
Lagrangian manifold is a priori not the eigenstate of a specific
model, but should be viewed as the possible initial state of some
quantum evolution to be specified later. The
Lagrangian manifolds that we consider here are $K$-dimensional
manifolds in the $2K$-dimensional phase space $\mathcal{M}$, that are
characterized by the existence of a function $S(\bq)$ such that the
$\bp$ coordinates can be expressed as the gradient of that function,
namely $\bp=\partial \Sq/\partial \bq$ (Theorem 4.20 of \cite{Maslov81}).

A natural way to construct a Lagrangian manifold from state
$\psi_0(\bq)=\cala(\bq) \exp(i \Sq(\bq)/\hbar)$ is to consider the set
$\LL$ of phase-space points such that $\bp=\partial
\Sq/\partial \bq$: by definition it is a Lagrangian
manifold. Conversely, for any Lagrangian manifold $\LL$,
and given any smooth function $\cala(\bq)$, one
can construct the semiclassical state  $\cala(\bq) \exp(i \Sq_\LL
(\bq)/\hbar)$ with $S_\LL(\bq) \equiv \int_{\gamma:\bx0\to \bbx}
\bp d \bq$, where $\gamma$ is any path on $\LL$ going from an
arbitrary $\bx0$ to $\bbx=(\bp,\bq)$. Indeed, a consequence of the fact
that $\LL$ is Lagrangian is that the integral defining
$\Sq_\LL(\bq)$ does not depend on the path of integration but
only on the endpoints. 

\subsection{One-dimensional case}

\begin{figure}[tb]
\begin{center}
\includegraphics[width=0.45\linewidth]{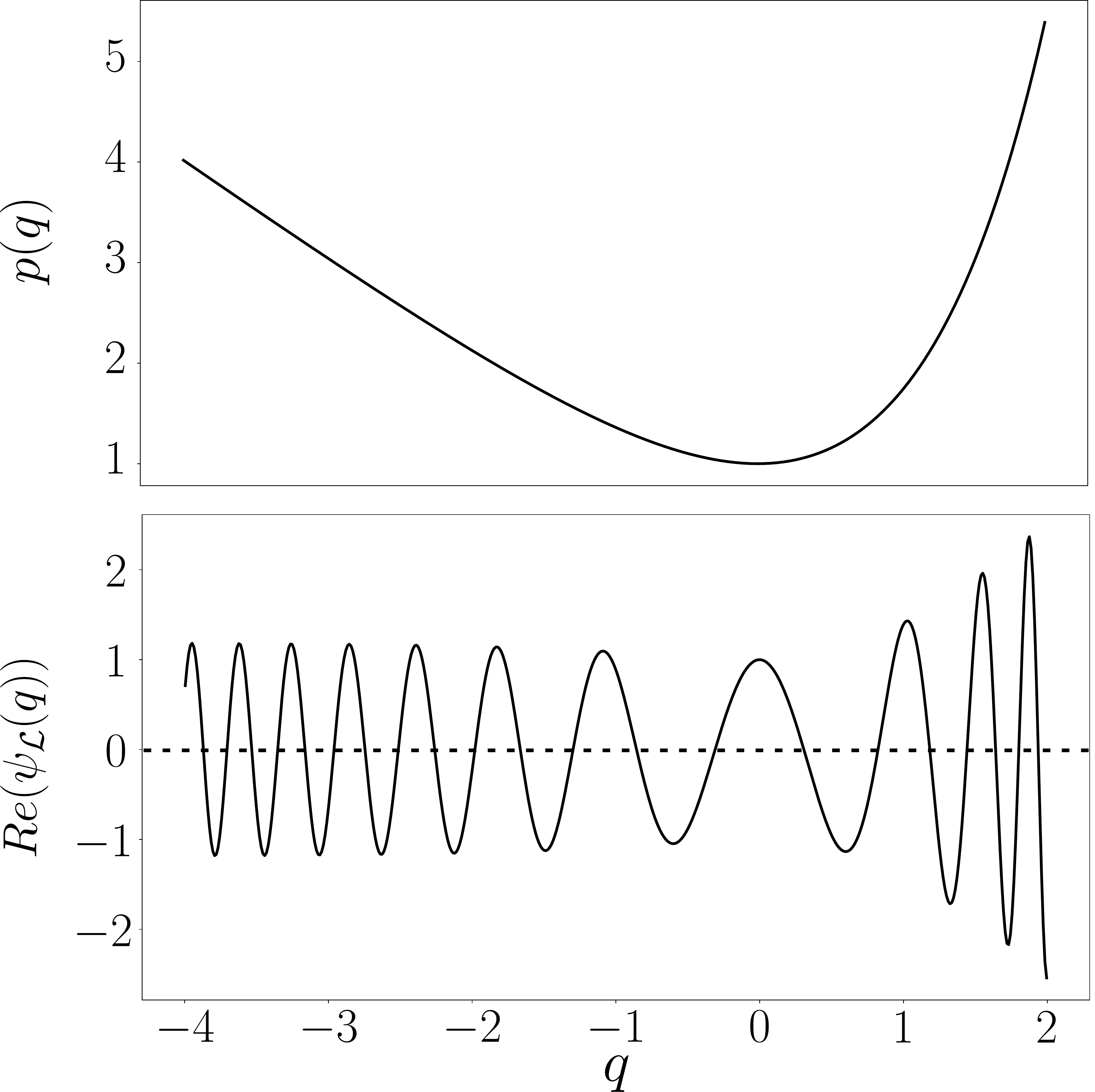}
\caption{Example of the association of a semiclassical wavefunction with a Lagrangian manifold. Top: the ``Lagrangian manifold'' (here, for $K=1$ just a curve) defined by $p(q) = e^q-q$. Bottom: real part of the associated wavefunction, assuming $a(s)=1$ with the parameter $s$ chosen as the curvilinear distance along the curve (i.e.~$ds^2 = dq^2 + dp^2$) and a value of $\hbar = 0.2$. \label{fig:maslov}}
\end{center}
\end{figure}

As an illustration, let us consider the one-dimensional case $K=1$. The phase space $\mathcal{M}$ is now two-dimensional, and a one-dimensional Lagrangian manifold $\LL$ is simply a curve in phase space, specified by some function $\Sq_\LL(q)$ as the set of points $(p,q)$ with $p=d\Sq_\LL(q)/dq$. A path $\gamma$ on $\LL$ is a portion of that curve connecting an initial point $x_0$ to a final point $x$, and the function $\Sq_\LL(q)$ can be expressed as $\Sq_\LL(q)  = \int_{\gamma : x_0 \to x} pdq$, with $x_0$ arbitrarily fixed on $\LL$. We parameterize the curve $\LL$ as $\LL = \{ x(s), s\in [0,1] \}$, with of course $x(s)=(p(s),q(s))$, and possibly $x(0)=x(1)$. We then associate with $\LL$ a semiclassical wavefunction following the procedure indicated above (see Fig.~\ref{fig:maslov} for an illustration).
Choosing a prefactor of the form $\cala(q)=a(s)/\sqrt{|dq/ds|}$ in a domain where $q(s)$ is monotonous (thus invertible), we set 
\begin{equation}
\label{eq:canonical_q} 
\psi_{\LL}(q)  = \frac{a(s)}{\sqrt{|dq/ds|}}
                            \exp\left(\frac{i}{\hbar} \Sq_\LL(q) - i
                        \frac{\pi}{2} \mu
                            \right), \qquad
\Sq_\LL(q)  = \int_{\gamma : x_0 \to x} p(s) \frac{dq}{ds} ds \; ,
\end{equation}
where $a:[0,1] \to \mathbb{C}$ is some smooth function of $s$, evaluated at the parameter value corresponding to $q$,  and the index $\mu$, introduced for later convenience, is a Maslov index which is zero in this simple case.
To go from $q$ to $p$ representation we define the Fourier transform of a function $\psi(q)$ as 
\begin{equation}
\label{ft}
\tilde\psi(p)=\frac{1}{(2i\pi\hbar)^{1/2}}\int dq\,e^{-\frac{i}{\hbar}pq}\psi(q).
\end{equation}
The Fourier transform \eqref{ft} has the property that $|\tilde\psi(p)|^2 |\partial p/\partial q|=|\psi(q)|^2 |\partial q/\partial p|$ up to $1/\hbar$ corrections. If we calculate the Fourier transform $\tilde\psi_{\LL}(p)$ of $\psi_{\LL}(q)$ in the stationary phase approximation, we get that the stationary point is at a value of $q$ such that $d\Sq_\LL/dq=p$, and Eq.~\eqref{ft} then gives
\begin{equation}
\label{eq:canonical_p}
\tilde\psi_{\LL}(p)
 = \frac{a(s)}{\sqrt{|dp/ds|}}\exp\left(\frac{i}{\hbar} \tilde \Sq_\LL(p) - i \frac{\pi}{2} \tilde \mu \right), \qquad
\tilde \Sq_\LL(p) =-\int_{\LL : x_0 \to x} q(s) \frac{dp}{ds} ds \; ,
\end{equation}
where $\tilde \mu=1$ if $dp/dq=d^2\Sq_\LL/dq^2$ is negative and 0 otherwise (Corollary 1.9 of \cite{Maslov81}). Here the prefactor $a(s)$ is evaluated at the parameter value $s$ corresponding to the stationary point $q$ at which $d\Sq_\LL/dq=p$, that is, at the parameter value $s$ corresponding to $p$ itself. Remarkably, because of the choice of the form of the prefactor $\cala(q)$, the semiclassical wavefunctions \eqref{eq:canonical_q} and \eqref{eq:canonical_p} have exactly the same form, up to the sign in front of the action. 

Expression \eqref{eq:canonical_p} is valid only in the case where $dp/ds\neq 0$. 
Of course, there is no special role played by variable $q$ in the above construction, and one can as well directly associate with $\LL$ a semiclassical wavefunction in the $p$ representation. Such a function can be obtained by changing $(p,q)$ to $(q, -p)$ in Eq.~\eqref{eq:canonical_q}, and it will be defined in a domain where $p(s)$ is invertible. In fact, this function exactly takes the form \eqref{eq:canonical_p}, and therefore we denote it by $\tilde\psi_{\LL}(p)$. Thus, the association between a manifold $\LL$ and a wavefunction $\psi_{\LL}$ has this nice property that it can be performed in a symmetric way in the $q$ or the $p$ representation. 

If both $q(s)$ and $p(s)$ are invertible, it is equivalent (up to a constant phase factor) to obtain the semiclassical wavefunction using either representation. For a generic $\LL$ however, neither $q(s)$ nor $p(s)$ will be invertible over the whole parameter range (think of an ellipse parameterized by an angle). The procedure suggested by Maslov is then to introduce
a partition of the unity, i.e.~a set of $n$ smooth positive real
functions $\{\varphi_\alpha(s),\alpha = 1,\ldots,n\}$ such that
$\sum_1^n \varphi_\alpha(s) = 1$, and to choose it such that for any
$\alpha$, either $q(s)$ or $p(s)$ is invertible on the
support of $\varphi_\alpha$. In that case one can define a function $\psi_\alpha(q)$ in $q$ representation using either Eq.~\eqref{eq:canonical_q} (if $q(s)$ is invertible) or the inverse Fourier transform of Eq.~\eqref{eq:canonical_p} (if $p(s)$ is invertible), in both cases with amplitude $\varphi_\alpha(s)a(s)$. The semiclassical wavefunction associated with $\LL$ is then simply $\psi_{\LL}(q) = \sum_\alpha \psi_\alpha(q)$. 
It can then be shown that if in
Eqs.~\eqref{eq:canonical_q}-\eqref{eq:canonical_p} the indices
$\mu_\alpha$, $\tilde \mu_\alpha$ count the (algebraic) number of
caustics from $x_0$ to the support of $\varphi_\alpha$ for the
corresponding representation, the resulting wavefunction is, up to
negligible higher order in $\hbar$ corrections, independent on the
details of the choice of the $\varphi_\alpha$, and in particular
that the phase is a continuous function.

If $\LL$ is not a closed curve, this completes the construction.  If $x(0)=x(1)$ however, one needs to further impose that the total phase is single-valued, which leads to the usual EBK quantization condition 
\begin{equation} \label{eq:EBK}
\oint_\LL pdq = (n+\mu_\LL/4) 2\pi\hbar \; 
\end{equation}
with $\mu_\LL$ the Maslov index associated with $\LL$
(which is usually 2 when $K=1$). We stress
again that we did not introduce any dynamics yet (i.e.~we did not
specify the Hamiltonian, which could for instance be a time-dependent
one), and the EBK condition \eqref{eq:EBK} does not specify an eigenstate of
the problem, but just a constraint on the manifold $\LL$ such
that one can associate to it a wavefunction using the Maslov
procedure.

\subsection{Generalization to arbitrary dimension}

This construction generalizes straightforwardly to the general case of
a $K$-dimensional system. The symplectic structure of phase space
$\calm$ can be introduced through the wedge product giving the
symplectic area of a pair of phase-space vectors $(\bbx,\bbx')$,
\begin{equation} \label{eq:wedge}
\bbx \wedge \bbx' = \sum_{i=1}^{K} \left(p_iq'_i-q_ip'_i\right)=
(\bbx')^T J \bbx,\qquad  \quad 
J=\left(\begin{array}{cc}0&-\idmat\\ \idmat&0\end{array}\right)
\end{equation}
where $\idmat$ is the $K\times K$ identity matrix. A Lagrangian manifold is then a manifold on
which the differential 2-form $\omega^{2} = \sum_{i=1}^K dp_i \wedge dq_i$ is
uniformly zero. As a consequence, the integral of the 1-form
$\bp d \bq \equiv \sum_{i=1}^K p_i dq_i$ along any path on
$\LL$ is invariant under a continuous deformation of that path with
  fixed endpoints.
  
 If $\LL$ is $K$-dimensional, we can
parameterize it by a vector of parameters
$\bs\in\mathbb{R}^K$. Formulas \eqref{eq:canonical_q}-\eqref{eq:EBK}
are easily adapted if we use the compact notation where
$\partial \bq/ \partial \bs$ denotes the $K\times K$ matrix with
matrix elements $\partial q_i/\partial s_j$ and
$\partial \Sq /\partial \bq$ denotes the vector with components
$\partial \Sq /\partial q_i$. Then the Lagrangian manifold
$\LL$ is specified by some function $S_\LL(\bq)$, and
\eqref{eq:canonical_q} becomes
\begin{equation}
\label{eq:canonical_q_K} 
\psi_{\LL}(\bq)  = \frac{a(\bs)}{\sqrt{|\partial \bq/\partial \bs|}}
                            \exp\left(\frac{i}{\hbar} \Sq_\LL(\bq) - i
                        \frac{\pi}{2} \mu
                            \right), \qquad
\Sq_\LL(\bq)  = \int_{\gamma : \bbx_0 \to \bbx} \bp(\bs) \frac{\partial\bq}{\partial\bs} d\bs \; ,
\end{equation}
where $|.|$ denotes the absolute value of the determinant of the matrix.
Since $\LL$ is Lagrangian, the
action integral in \eqref{eq:canonical_q_K} is invariant under a continuous deformation of the path of
integration on $\LL$ and thus, once EBK conditions such as
\eqref{eq:EBK} are fulfilled for all generators of the homotopy group
of $\LL$, phase factors only depend on the point
$\bbx\in\LL$ but not on the path chosen to define the action
integral.  

At each point a $K$-dimensional
Lagrangian manifold $\LL$ is locally diffeomorphic to a
coordinate $K$-dimensional plane (the $\bq$ plane, the $\bp$ plane, or
a mixture of both) (Proposition 4.6 of \cite{Maslov81}), which allows to always
choose local coordinates such that the parametrization is
non-singular, so that one can always construct either the
semiclassical wavefunction \eqref{eq:canonical_q}, or
\eqref{eq:canonical_p}, or obtain an analog expression in a mixed
representation. 


\section{Semiclassical Wigner function}
\label{scwf}

In this section, we derive a semiclassical expression for the Wigner
function, which will be the starting point of our discussion in
section~\ref{sec:expectation}.  We follow here very closely the spirit
of the original work of Berry \cite{Ber77}, although the precise
expressions we shall obtain will take a slightly different form, whose
transformation under {\em canonical changes of variables} will be
somewhat more transparent.

\subsection{Stationary phase approximation}
\label{wignerstatphase}
The Wigner transform $W_\LL(\bbx)$ of a semiclassical wavefunction 
$\psi_{\LL}$ of the form \eqref{eq:canonical_q_K} is obtained
using the definition \eqref{eq:Wignerpsi}. In order to obtain the
semiclassical expression for $W_{\LL}(\bbx)$, we need to
evaluate the integral   
\begin{equation} 
\label{eq:WignerpsiK}
W_{\LL}(\bbx)=\frac{1}{(2\pi \hbar)^K}\int \dK{\bq'} e^{-\frac{i}{\hbar}\bp.\bq'}\overline{\cala(\bq-\frac{\bq'}{2})}\cala(\bq+\frac{\bq'}{2})
e^{\frac{i}{\hbar} \Sq_\LL(\bq+\frac{\bq'}{2})- \frac{i}{\hbar} \Sq_\LL(\bq-\frac{\bq'}{2})-i\frac{\pi}{2}(\mu_{\bq+\bq'/2}-\mu_{\bq-\bq'/2})}
\end{equation}
in the stationary phase approximation \cite{Ber77}.

At a given $\bbx=(\bp,\bq)$, a stationary point in the integral of Eq.~\eqref{eq:WignerpsiK} is reached when $\bq'$ takes a value $\tilde{\bq}$ such that $\bp = \frac12 [\partial \Sq_\LL /\partial   \bq|_{\bq-\tilde{\bq}/2} + \partial \Sq_\LL / \partial   \bq|_{\bq+\tilde{\bq}/2} ]$. Symmetrically, a stationary point is also reached at $\bq'=-\tilde{\bq}$. Let $\bq^+\equiv \bq + \tilde{\bq}/2$. According to the above definitions, $\bp^+=(\partial \Sq_\LL /\partial \bq)_{|\bq^+}$ defines a point $\bbx^+=(\bp^+,\bq^+)$ on $\LL$, corresponding to a parameter value $\bs_+$. Similarly, if $\bq^-\equiv \bq - \tilde{\bq}/2$, then $\bp^-=(\partial \Sq_\LL /\partial \bq)_{|\bq^-}$ defines a point  $\bbx^-=(\bp^-,\bq^-)$ on $\LL$, corresponding to a parameter value $\bs_-$. The stationary condition can then be rewritten as $\bp=\frac12(\bp^++\bp^-)$. Since obviously $ \frac12 (\bq^++\bq^-)=\bq$, we have that $\bbx = \frac12 (\bbx^+ + \bbx^-)$. Therefore, for a stationary point to exist in the integral evaluating $W_\LL(\bbx)$, the point $\bbx$ has to be the {\em center} of a pair of points $(\bbx^+,\bbx^-)$ lying on $\LL$.  We shall below refer to $\bxi  =  (\bp^+-\bp^-,\bq^+-\bq^-)$ as the corresponding {\em chord}.

\begin{figure}[tb]
\begin{center}
\includegraphics[width=0.45\linewidth]{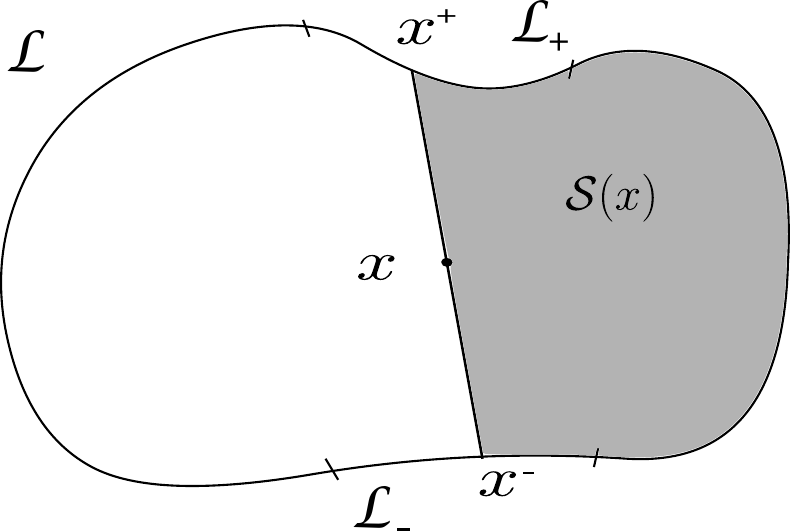}
\caption{Construction of the action $\Sx(\bbx)$ introduced in Eq.~\eqref{Sxbx}.
For a given $\bbx$, one needs to locate two points $\bbx^+$ and $\bbx^-$ on the manifold $\LL$ such that $\bbx$ is the center of the segment $[\bbx^-,\bbx^+]$ (more than one pair of points may fulfill that condition; such pairs are indexed by $\beta$ in the text).  The action $\Sx(\bbx)$ then corresponds to the shaded area (note that because of the EBK quantization condition Eq.~\eqref{eq:EBK}, which side of the chord is used is irrelevant).  
$\LL_+$ and $\LL_-$ are pieces of $\LL$ on which the endpoints $\bbx^+$ and $\bbx^-$ lie.  
\label{fig:wigner}}
\end{center}
\end{figure}

We find it convenient to denote with superscripts $\pm$ all quantities evaluated at point $\bq^+$ or $\bq^-$ on $\LL$, parameterized by the parameter value $\bs^\pm$. All quantities at the stationary point can then be expressed as functions of $\bq^+$ and $\bq^-$, or equivalently of $\bs^+$ and $\bs^-$. 
At the stationary point $\bq'=\tilde{\bq}$, the phase of the integrand in Eq.~\eqref{eq:WignerpsiK} is $\Sx(\bbx) / \hbar$ with
\begin{equation}
\label{Sxbx}
\Sx(\bbx)  = \Sq_\LL(\bq^+) - \Sq_\LL(\bq^-) - \bp\tilde{\bq}=  \oint_{\bbx^-}^{\bbx^+}\bp(\bs) \frac{\partial\bq}{\partial\bs} d\bs \; ,
\end{equation}
where in the right-hand side the integral is taken along the closed contour made of any path going from $\bbx^-$ to $\bbx^+$ on $\LL$ and the straight line from $\bbx^+$ back to $\bbx^-$ (see Fig.~\ref{fig:wigner}). The Hessian matrix $\partial^2 \Sq_\LL/\partial \bq^2=\partial  \bp/\partial \bq$, evaluated at $\bq^+$, is  denoted by $\partial \bp^+/\partial \bq^+$. We define the matrix
\begin{equation}
\label{uneq}
R\equiv \frac{\partial^2 \Sq_\LL}{\partial \bq^2}\Big|_{\bq^+}
- \frac{\partial^2 \Sq_\LL}{\partial \bq^2}\Big|_{\bq^-} = \left( \frac{\partial \bp^+}{\partial\bq^+}  \right)^T - \left(\frac{\partial \bp^-}{\partial \bq^-}\right)
\end{equation}
(since the Hessian is a symmetric matrix, we may or may not take the transpose in the right-hand side of \eqref{uneq}). Let $\eta$ denote the number of negative eigenvalues of $R$. At the stationary point, Gaussian integration yields 
\begin{equation} 
  W_{\LL}(\bbx)  = \left(\frac{2}{\pi\hbar}\right)^{K/2} \sum_{\beta}\frac{\cala(\bq^+_{\beta})\cala(\bq^-_{\beta})}{\sqrt{\left|R_\beta \right|}} \exp \left( \frac{i}{\hbar}\Sx_{\beta}(\bbx) - i \frac{\pi}{2} \mu_{\beta} +i\eta_{\beta} \frac{\pi}{4}\right)+\textrm{ c.c.}\;,  \label{eq:SC-Wigneraux}  
\end{equation}
where the complex conjugate c.c.~corresponds to the contribution of $\bq'=-\tilde{\bq}$ \cite{Ber77}.
In \eqref{eq:SC-Wigneraux} the sum runs over all stationary phase points $\pm(\tilde{\bq})_{\beta}$,
that is, all {\em chords} $\bxi_{\beta}$ such that the two points
$\bbx_{\beta}^- = \bbx -\bxi_{\beta}/2$ and $\bbx_{\beta}^+ = \bbx
+\bxi_{\beta}/2$ associated with the {\em center} $\bbx$ and the chord
$\bxi_{\beta}$ lie on $\LL$. In Eq.~\eqref{eq:SC-Wigner},
$\mu_{\beta}=\mu_{\beta}^+-\mu_{\beta}^-$ is the Maslov index computed along the path on $\LL$
from $\bbx_{\beta}^-$ to $\bbx_{\beta}^+$. The phase $\Sx_{\beta}(\bbx)$ can be interpreted as the symplectic area
enclosed by the path from $\bbx_{\beta}^-$ to $\bbx_{\beta}^+$ and the
chord $\bxi_{\beta}$. As a consequence, its variation $\delta \Sx_\beta$ when the phase point $\bbx$ is moved by $\delta\bbx$ is, at first order in this small quantity, the
symplectic area of the quadrilateral formed by the chord $\bxi_{\beta}$
and the displacement $\delta\bbx$. We thus have 
$\delta\Sx_\beta = \bxi_{\beta} \wedge \delta\bbx =
(\delta\bbx)^T.J\bxi_{\beta}$, and  therefore
\begin{equation} 
\label{eq:dSLdx}
\frac{\partial \Sx_{\beta}}{\partial \bbx} = J \bxi_{\beta} \; .
\end{equation}

\subsection{Canonical invariance of the Wigner function}
\label{sec:canonical}

The ``exact'' quantum-mechanical Wigner function is invariant under
{\em symplectic maps}, that is {\em linear canonical
    transformations} \cite{Voros77, OzoHan82}. However this is not
  entirely visible from the definition Eq.~\eqref{eq:Wignerpsi}, and
  is often not very transparent either for semiclassical
  approximations of the Wigner function (see however \cite{OzoHan82},
  in which the amplitude of the semiclassical Wigner function is
  expressed in terms of Poisson brackets of a pair of actions).  In
  this subsection we will cast \eqref{eq:SC-Wigneraux} in a form which
  is manifestly invariant under linear canonical transformations.

Indeed, using the fact that
  $\cala(\bq)=a(\bs)/\sqrt{|\partial\bq/\partial\bs|}$, the
  denominator in \eqref{eq:SC-Wigneraux} can be written in
    terms of the $K\times K$ matrix $D$ defined as
\begin{equation}
\label{matrixD}
D\equiv\left(\frac{\partial \bq^+}{\partial \bs^+} \right)^T 
\left [\left(\frac{\partial
    \bp^+}{\partial \bq^+}\right)^T - \left(\frac{\partial
    \bp^-}{\partial \bq^-}\right)  \right] \left(\frac{\partial
\bq^-}{\partial \bs^-} \right)
=\left(\frac{\partial
    \bp^+}{\partial \bs^+}\right)^T   \left(\frac{\partial
\bq^-}{\partial \bs^-}\right) - 
\left(\frac{\partial
    \bq^+}{\partial \bs^+}\right)^T   \left(\frac{\partial
\bp^-}{\partial \bs^-}\right) \; ,
\end{equation}
and which components are given by
\begin{equation}
\label{eq:Dab}
D_{ab}= \frac{\partial \bbx^+}{\partial s^+_a}\wedge \frac{\partial
  \bbx^-}{\partial s^-_b} \; .
\end{equation}
The semiclassical Wigner function can then be rewritten as
\begin{equation} \label{eq:SC-Wigner}
  W_{\LL}(\bbx) =   \left(\frac{2}{\pi\hbar}\right)^{K/2} \sum_{\beta}
  \frac{\overline{a(\bs_{\beta}^-)}
    a(\bs_{\beta}^+)}{\sqrt{\left|D^\beta \right|}} \exp
  \left(\frac{i}{\hbar} \Sx_{\beta}(\bbx) - i \frac{\pi}{2}
    \mu_{\beta} +i\eta_{\beta} \frac{\pi}{4}\right)+\textrm{ c.c.} \; .
\end{equation}
Under the form \eqref{eq:SC-Wigner} it becomes clear that
$W_{\LL}(\bbx)$ is invariant under symplectic maps. Indeed, such transformations preserve the
wedge product, so that all $D_{ab}$ are invariant. As for the action
$\Sx_{\beta}(\bbx)$, it is given by the symplectic area \eqref{Sxbx}
enclosed by a path drawn on the Lagrangian manifold together with the
chord joining the two points $\bbx^+$ and $\bbx^-$. The symplectic
area enclosed by a given curve is invariant under any  canonical
transformation; and for a symplectic map, the chord joining $\bbx^+$
to $\bbx^-$ is transformed into the chord joining the image of $\bbx^+$
and $\bbx^-$. Thus, \eqref{eq:SC-Wigner} is invariant under symplectic maps.

However, $W_{\LL}(\bbx)$ is {\em a priori} not
  necessarily invariant under {\em generic} canonical transformations.
  Consider first the matrix $D$.  Canonical transformations can be
  characterized by the fact that they preserve the 2-form $\omega^2$
  derived from the skew product \eqref{eq:wedge}. This 2-form is
  defined at each point on the local tangent space, but
  $\partial \bbx^+/\partial s^+_a$ and
  $\partial \bbx^-/\partial s^-_b$ in \eqref{eq:Dab} should be
  considered as belonging to the local tangent space at $\bbx^+$ and
  $\bbx^-$ respectively. They will transform as a vector field under
a canonical transformation, i.e.~along the corresponding linearized
maps $M_+$ and $M_-$ respectively, so that
\begin{equation}
D_{ab}=\frac{\partial
  \bbx^-}{\partial s^-_b} J \frac{\partial \bbx^+}{\partial s^+_a}\to
\frac{\partial
  \bbx^-}{\partial s^-_b}M_-^T J M_+ \frac{\partial \bbx^+}{\partial s^+_a}.
\end{equation}
Since in general $M_-^T J M_+\neq J$ the quantities $D_{ab}$ are not
invariant (of course for symplectic maps $M_+=M_-$ and we recover
invariance). The same goes for the action $\Sx_{\beta}(\bbx)$
  since for a generic canonical transformation the chord joining
  $\bbx^+$ and $\bbx^-$ does not {\em a priori} transform into a
  chord.

  This statement being made, we shall however see in
  section~\ref{sec:expectation} that the interference contributions to
  the mean value Eq.~\eqref{hata} are dominated by {\em short chords},
  i.e.\ chords $\bxi_{\beta} =\bbx^+_{\beta} -\bbx^-_{\beta}$ whose
  length goes to zero in the semiclassical limit.  This means that as
  $\hbar \to 0$, the contributions of the Wigner function we will be
  interested in will be the ones such that
  $ \bbx^+_{\beta} \to \bbx^-_{\beta}$ (although
  $\bs^+_{\beta} \not \to \bs^-_{\beta}$), see Fig.~\ref{fig:coordinate}. For these contributions,
  the distinction between general and linear canonical maps becomes
  irrelevant. Indeed the tangent spaces near $ \bbx^+_{\beta}$
    and $ \bbx^-_{\beta}$, as well as the linearized maps $M_+$ and
    $M_-$, can be identified, while the distortion of the chord between
    $ \bbx^+_{\beta}$ and $ \bbx^-_{\beta}$ becomes negligible. In
  that limit, both the matrix $D^\beta$ and the action
  $\Sx_{\beta}(\bbx)$ are genuine canonical invariants up to
  negligible higher-order $\hbar$ corrections.  Thus, the
    contributions $\beta$ which, in the sum \eqref{eq:SC-Wigner}, {\em
      are relevant for the calculation of the mean value of smooth
      operators} can be considered as invariant under any (i.e.\ not
    necessarily linear) canonical transformation.  Interestingly,
  going back from \eqref{eq:SC-Wigner} to \eqref{eq:WignerpsiK}, this
  also means that the same considerations can be applied to this
  original form of the Wigner function: in the semiclassical limit
  where only the vicinity of short chords contributes, one can choose
  arbitrary canonical coordinates to express \eqref{eq:WignerpsiK}, as
  all discrepancies will be higher-order $\hbar$ corrections. We will
  make use of this property in the following section, where we shall
  allow ourselves arbitrary canonical transformations, keeping in mind
  that this only makes sense in the situation where short chords
  dominate.

\subsection{Parametrization invariance of the Wigner function}\label{invpar}
The amplitude of
$\psi_{\LL}(\bq)$ in \eqref{eq:canonical_q_K} is $\cala(\bq)$,
which should  not depend on the parametrisation $\bs$ of the Lagrangian manifold. The Wigner function in \eqref{eq:WignerpsiK} is therefore invariant under a change of parameter, and so should be its semiclassical expression \eqref{eq:SC-Wigner}. This can be checked directly on \eqref{eq:SC-Wigner}. Indeed, upon a change
$\bs\mapsto\bs'$, $a(\bs)$ should transform as
$\frac{a(\bs)}{\sqrt{|\partial \bq/\partial
    \bs|}}=\frac{a(\bs')}{\sqrt{|\partial \bq/\partial \bs'|}}$,
i.e.
\begin{align}
\bs & \mapsto \bs' \\
a(\bs) & \mapsto a(\bs')=a(\bs)\sqrt{|\partial \bs/\partial \bs'|} \; . \label{eq:stosp}
\end{align}
Since using \eqref{eq:Dab} we have 
\begin{equation}
\label{transformd}
D_{ab}(\bs) \mapsto D_{ab}(\bs') 
=\sum_{i,j=1}^K\frac{\partial s_i}{\partial s_a'}\left(\frac{\partial
    \bbx^+}{\partial s^+_i}\wedge \frac{\partial \bbx^-}{\partial
    s^-_j}\right)\frac{\partial s_j}{\partial s_b'} \; ,
\end{equation}
the ratio $\overline{a(\bs^{-})}a(\bs^{+})/\sqrt{|D|}$ is as expected invariant.
The choice of parameter will be guided by convenience. For instance, when considering the eigenstates of classically integrable systems, for which one can define a system of action-angle coordinates, the Lagrangian manifolds of interest are the invariant tori specified by the action variables, and the parameters $\bs$ can be chosen as the angle variables \cite{OzoHan82}.
In Section \ref{sec:expectation} we will choose a local parametrization that simplifies calculations for mean values of observables.

\subsection{Wigner function in the vicinity of the manifold}
The above semiclassical expansion \eqref{eq:SC-Wigner} is not valid when
$\bbx$ is in the vicinity of $\LL$. Indeed, in the limit $\bbx\to\LL$ we
have $\bs^+\to \bs^-$ and $\bbx^+\to\bbx^-$, and therefore the matrix
$D$ becomes singular. The stationary approximation is no longer valid in this case, and we have to start over from \eqref{eq:WignerpsiK}.

Let us assume that $\LL$ has some curvature only at a classical scale. Keeping only linear terms in the actions $\Sq_\LL$ in \eqref{eq:WignerpsiK}, we get (disregarding the constant phases)
\begin{equation} 
\label{intdiagonal}
W_{\LL}(\bbx)=\frac{1}{(2\pi \hbar)^K}\int \dK{\bq'} \overline{\cala(\bq-\frac{\bq'}{2})}\cala(\bq+\frac{\bq'}{2})e^{\frac{i}{\hbar}(\bp_\LL-\bp)\bq'} \; ,\quad\textrm{with } \bp_\LL=\frac{\partial \Sq_\LL}{\partial  \bq}.
\end{equation}
If the prefactor varies slowly as a function of its arguments, that is, only on a classical scale, then at $\hbar\to 0$ the amplitudes can be taken out of the integral, so that \cite{Heller77}
\begin{equation} 
\label{Wdiag}
W_{\LL}(\bbx) = |\cala(\bq)|^2\delta(\bp-\bp_\LL) 
=\int \dK{\bq_\LL} |\cala(\bq_\LL)|^2\delta(\bbx-\bbx_\LL) 
= \int \dK{\bs} |a(\bs)|^2 \delta(\bbx - \bbx(\bs))\; ,
\end{equation}
where the last equality comes from the change of variables $\bq_\LL\mapsto\bs$, using \eqref{eq:stosp}. The right-hand side of \eqref{Wdiag} is again manifestly canonically invariant since, upon integration over an aribtrary test function it will only depend on geometrical properties of $\LL$.
\rem{citer Heller 77}

At small but finite $\hbar$, taking into account the curvature of the
manifold through uniform approximation will lead to an Airy function
in the one-dimensional case \cite{Ber77, BerBal78}, and to more subtle
behaviors in higher dimension \cite{OzoHan82}. Taking into account
such corrections, or the the variation of the prefactor, will in
practice broaden the $\delta$ function on the quantum scale. If the
Wigner function is convoluted with a classical object, as we shall do
below, this will however not affect the final result.

\subsection{Time propagation}
\label{sec:TimeProp}

We now introduce a dynamics through some Hamiltonian $\hat H$.
Let $\gt$ be the corresponding classical Hamiltonian flow. For any fixed $t$ it is a canonical transformation (Theorem 4.17 of \cite{Maslov81}). Since any canonical transformation  maps a Lagrangian
manifold into another one (Theorem 4.18 of \cite{Maslov81}), the manifold $\LL_t=\gt
\LL$ obtained by propagation of $\LL$ under the classical flow is a  Lagrangian
manifold. 

The strength and beauty of the Maslov association between the Lagrangian
manifold $\LL$ and the semiclassical wavefunction
$\psi_\LL$ is that the semiclassical propagated wavefunction 
$\psi_\LL(t)$ is obtained from the manifold $\LL_t=\gt
\LL$ following exactly the same procedure. Therefore, the time evolution of the Wigner function  
$W_{\LL}(\bbx)$ is obtained as 
\begin{equation} \label{eq:SC-Wigner-prop} 
W_{\LL}(\bbx,t) = W_{\LL_t}(\bbx),\quad \LL_t=\gt\LL\,.
\end{equation}
This means that the Wigner function at time $t$ is given by Eq.~\eqref{eq:SC-Wigner} with $\LL$ replaced by its time evolution $\gt\LL$  (and coordinates $\bs$ on $\gt \LL$ are the image of the coordinates $\bs$ on $\LL$). Equation \eqref{eq:SC-Wigner} and its time evolved version
Eq.~\eqref{eq:SC-Wigner-prop} will form the basis of our discussion of
the mean value of smooth operators.

\section{Expectation value of smooth operators}
\label{sec:expectation}

\subsection{Stationary point contribution, and why this is not enough}

We now turn to the discussion of the time evolution of the expectation
value of an operator $\hatO$ for an initial semiclassical
wavefunction $\Psi_0$ constructed on an initial manifold $\LL_0$
following the Maslov approach described in section~\ref{sec:SemiclassicWF}. 
From the discussion of section~\ref{sec:TimeProp}, the Wigner function $W_t(\bbx)$ of the time-evolved wavefunction is given by $W_{\LL_t}(\bbx)$, expressed as Eq.~\eqref{eq:SC-Wigner}, with $\LL_t=\gt \LL_0$. The expectation value is then given by Eq.~\eqref{hata}. 

We will assume that $\hat O$ is a {\em classical} operator, in the
sense that its Wigner transform $O(\bbx)$ behaves like a classical
quantity, that is, it has no significant variation on the quantum
scale fixed by $\hbar$.  This notion that $\hatO$ is classical implies
that, in the stationary phase approximation of \eqref{hata}, only the
phase of $W_t(\bbx)$ has to be considered. This phase, appearing in
\eqref{eq:SC-Wigner}, is given by \eqref{Sxbx} in which the action
integrals are taken on the propagated manifold $\LL_t = \gt \LL_0$.
Its derivative with respect to the integration variable $\bbx$ is
given by Eq.~\eqref{eq:dSLdx}, namely
$\partial\Sx_\beta(\bbx)/\partial \bbx =J \bxi$, where $\bxi$ is the
chord joining two points on $\LL_t$ and having $\bbx$ as center. Thus
the stationary phase condition is given by $\bxi=0$.  This implies
that $\bbx \in \LL_t$. The semiclassical expression
\eqref{eq:SC-Wigner} is actually not valid there, as the prefactor
$1/\sqrt{|D^\beta|}$ diverges. Nevertheless, if one starts from the
original expression Eq.~\eqref{eq:WignerpsiK}, the stationary phase
condition indeed corresponds to chords of length zero.

This implies obviously that there is no hope to find stationary phase
points outside of $\LL_t$, but also that there are $K$ directions
(corresponding to the tangents of $\LL_t$ at $\bbx$) for which $\bxi$
remains zero, and thus half of the eigenvalues of
$(\partial\bxi/ \partial \bbx)$ are zero. As a consequence, one cannot do a simple Gaussian integral to take into account the contribution to \eqref{hata} of the neighborhood of the stationary phase points.  However, in the neighborhood of the manifold one can use the expression \eqref{Wdiag} for $W_{\LL_t}(\bbx)$, and the integral \eqref{hata} then gives
\begin{equation}
\label{eq:expectation}
\langle\hatO\rangle_{\rm cl}(t)
=  \int_{\LL_t} \dK{\bs} \, O(\bbx(\bs))|a(\bs)|^2 \; ,
\end{equation}
which is interpreted as the classical contribution to the expectation
value, again in an explicitly canonically invariant form.  
Note here that if we accept the form \eqref{eq:expectation} at
  $t=0$ for $\langle\hatO\rangle_{\rm cl}$, application of
  \eqref{eq:TWA}  immediately yields the form \eqref{eq:expectation}
  for all times.  The classical contribution \eqref{eq:expectation}
  thus has  in some sense the same physical content as the TWA/SCL-IVR approximation \eqref{eq:TWA}.

The fact that at stationary points half of
the eigenvalues of $(\partial\bxi/ \partial \bbx)$ are zero is a sign
that the neighborhoods of stationary phase points are rather atypical,
which is presumably a first hint of why another kind of contribution
needs to be kept here.

Indeed, the essence of the stationary phase approximation is not so
much that only stationary phase points contribute (they usually are
of measure zero), but that the neighborhood of stationary phase
points contains all the points where the phase varies slowly.  Indeed,
if one considers the integral
\begin{equation} \label{eq:integral}
\int_{y_1}^{y_2} dy\, g(y) \exp \left[ \frac{i}{\hbar} f(y) \right] 
\end{equation}
in the small-$\hbar$ limit, and finds a point $y^*$ where $f$ varies
slowly, i.e.~$f'$ is small, then there should be a point in the
vicinity of $y^*$ where $f'$ vanishes. More precisely, if $y^*$ is
such that $f'(y^*) = \epsilon^* \ll \hbar (y_2 - y_1)$, then we can
expect that $y_{\rm st} \simeq y^* - \epsilon^* / f''$, with
$f'' = f''(y^*) \simeq f''(y_{\rm st})$, is a genuine stationary phase
point. Usually, $f'(y)$ and $f''(y)$ are uncorrelated quantities; and
if we vary $y$ until we find a point $y^*$ such that $f'(y^*)$ is
small, there is no reason a priori to expect that at this point
$f''(y^*)$ is small as well.  Thus a priori any point $y^*$ where
$f'(y^*)$ is small should be in the immediate neighborhood of a
stationary point $y_{\rm st}$ where $f'(y_{\rm st})=0$. Therefore its
contribution to \eqref{eq:integral} is taken into account as soon as
the contribution of {\em the neighborhood} of $y_{\rm st}$ is taken
into account (which is what is done in the stationary phase
approximation).

In  generic cases, the only way to fail to have all the ``nearly
stationary'' points in the neighborhood of an exactly stationary one is to
tune a parameter to set precisely to zero the second derivative of the
phase.  This situation can however be handled by uniform
approximation, and can actually be interpreted by saying that the
stationary phase point has moved to the complex phase space, but remains
nevertheless close.

The situation we consider is significantly different.  Indeed here
``almost stationary points'' are associated with small chords
$\bxi = \bbx^+(\bs^+) -\bbx^-(\bs^-)$.  The neighborhood of stationary
phase points corresponds to the situation where $\bs^+ \simeq \bs^-$.
However it may be that $\bbx^+\simeq \bbx^-$ while $\bs^+ \not \simeq \bs^-$: this situation arises in particular in the case where the endpoints $\bbx^+$
and $\bbx^-$ of the chord correspond to two distinct regions of the
evolved manifold $\LL_t = \gt \LL_0$. In such a case the geometry of the problem
forbids the presence of a neighboring stationary phase point (i.e.~with exactly $\bxi=0$). In the
following, we reserve the terminology ``small chord'' to the ones such
that $\bs^+ \not \simeq \bs^-$. The dominant interference
contributions to $\langle\hatO\rangle(t)$ will come from these small
chords $\bxi$.

The question we may ask is how much we should expect such small chords
to actually be present in phase space for the evolved manifold
$\LL_t$. It is clear for instance that if $\LL_0$ does not 
specifically display such short chords, we are not expecting them
either for rather short time where they should remain non-generic
features.  However, as time increases, the manifold $\LL_t$ will
generically expand (linearly for integrable systems, but exponentially
quickly for chaotic systems), which for bounded system, for which the
total phase space volume that can be explored is finite, implies that
the typical ``distance'' between close but distinct sheets of $\LL_t$
has to go to zero as $t\to\infty$. \rem{Berry tendrils}  Therefore, assuming a small {\em
  but fixed} value of $\hbar$, there should generically be a time at
which {\em interference effects associated with short chords} will set
in.

Let us illustrate this for instance with the case of a one dimensional
$(K=1)$ chaotic system characterized by a Lyapunov parameter
$\lambda$.  In that case we expect that the length of $\LL_t$ behaves
as $l_0 \exp[\lambda t]$, with $l_0$ the length of $\LL_0$.  If the
total accessible phase volume $\Vol$ (that is, the phase-space domain
inside which $W_t(\bbx)$ is confined) is finite, the typical separation
between neighboring sheets of $\LL_t$ should be
$O\left( (\Vol/l_0)\exp[-\lambda t] \right)$.  Small chords, and thus
interference effects, will typically set in for times of the order of
\begin{equation} \label{eq:tsc}
t_{\rm s.c.} = \frac{1}{\lambda} \log \left( \frac{\Vol l}{\hbar l_0} \right)
\; 
\end{equation}
with $l$ a typical length scale characterizing the Wigner transform $O(\bbx)$ of the observable $\hat O$.  The time scale $t_{\rm s.c.} $ is thus
essentially the Ehrenfest time for which the ``characteristic action''
used is the ratio ${\Vol l}/{\hbar l_0}$.  Being logarithmic in
$\hbar$, it is thus a fairly short time.

We  therefore consider the contribution to \eqref{hata} associated with
two portions of $\LL_t$, denoted by $\LL_+$ and $\LL_-$ and locally
parameterized by $\bs^{+}$ and $\bs^{-}$ respectively, which come close to
one another.  The corresponding
interference contribution to the expectation value 
is an integral of the form
\begin{equation}
\label{integralesimple}
\int \dK{\bbx} \cA(\bbx) e^{\frac{i}{\hbar} \Sx(\bbx)} \, ,
\end{equation}
where, dropping indices $f$,
$\beta$ and $\LL$,  the action $\Sx$ is given by \eqref{Sxbx} and the smoothly
varying prefactor by 
\begin{equation} \label{eq:def-cA}
\cA(\bbx) \equiv \left(\frac{2}{\pi\hbar}\right)^{K/2}\frac{\overline{a(\bs^{-})}a(\bs^{+})
e^{i \eta \pi/4-i\mu\pi/2}}{\sqrt{\left| D \right|}}O(\bbx) ,
\end{equation}
with $\bs^{+}$, $\bs^{-}$, $D$, $\eta$ and $\mu$ associated with the different
chords $\bxi$ joining $\LL_+$ and $\LL_-$ and going through $\bbx$, as in Eq.~\eqref{eq:SC-Wigner}.

\subsection{One-dimensional case}
\label{sec:one-d}

Our goal in this subsection is to perform the phase space
integral in Eq.~\eqref{integralesimple} in
the one-dimensional case $K=1$.  We place ourselves in the context of a chaotic system, where for time $t$ larger than the characteristic time $t_{\rm s.c.}$ defined in \eqref{eq:tsc}, the various sheets of the manifold $\LL_t$ will tend to align along the  manifold, and therefore locally be oriented along essentially the same direction. In this picture, there will be a ``parallel'' direction (along the general direction of both considered sheets of $\LL_t$) along which quantities vary only on a classical scale, and a ``transverse'' one, where variations are on the quantum scale (see Fig.~\ref{fig:coordinate} for illustration).

We shall see that the integration in the transverse direction, which is associated with rapid variations of the phase away from the middle of the two sheets, can be done analytically using stationary phase approximation. The stationary phase approximation on the other hand cannot be applied for structural reasons to the integral along the parallel direction, which corresponds to a much weaker (almost the absence of) variation of the phase and is much more idiosyncratic, and we shall discuss a couple of limiting cases.

\subsubsection{Canonical change of variables}

\begin{figure}[t!]
\begin{center}
\includegraphics[width=0.65\linewidth]{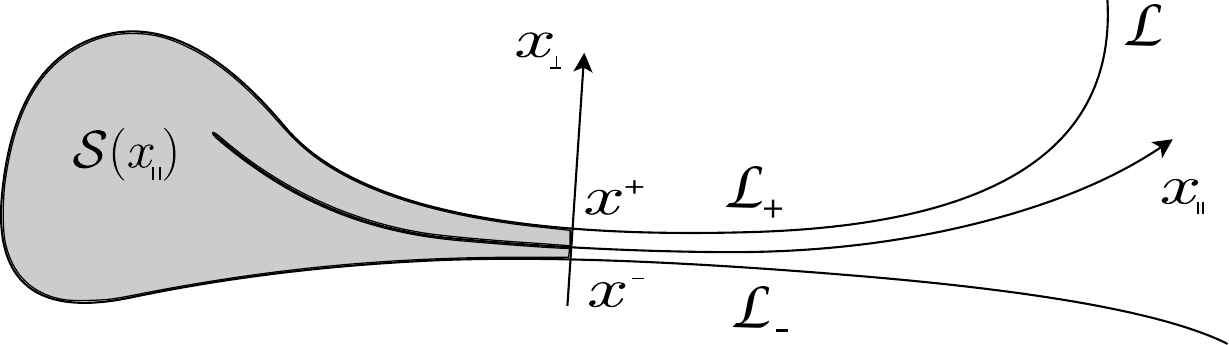}
\caption{Local system of coordinates used for evaluation of Eq.~\eqref{integralesimple} for small chords. The coordinate $\xpar$ is the parallel direction which follow the (almost identical on the classical scale) pieces $\LL_+$ and $\LL_-$ of the Lagrangian manifold $\LL$; $\xper$ is  the transverse direction. 
\label{fig:coordinate}}
\end{center}
\end{figure}

As discussed in section~\ref{sec:canonical}, when, as is the   case here, only {\em short chords} contributions are taken into   account, the expression of the semiclassical Wigner function
  \eqref{eq:SC-Wigner}, and thus the integral \eqref{integralesimple},
  is invariant under an arbitrary canonical
  change of variable. We therefore  introduce a pair of canonical variables $x = (\xpar,\xper)$ (see Fig.~\ref{fig:coordinate}), for
which the integral \eqref{integralesimple} simplifies. To do this, let
$\LLc$ be any smooth curve lying inbetween $\LL_+$ and $\LL_-$, and
$\lambda_0$ some coordinate on $\LL_c$. We note $x_0 = x(\lambda_0)$
the phase point on $\LL_c$ with coordinate $\lambda_0$. At each point
$x_0 \in\LLc$ the corresponding chord $\xi(x_0)$ defines a direction
in phase space. Any point $x$ in the region inbetween $\LL_+$ and
$\LL_-$ is of the form $x_0 + \abscisse \xi(x_0)$ for some $x_0 \in
\LLc$ and $\abscisse \in[-\frac12,\frac12]$. One can therefore adopt a
local coordinate system such that $x=(\xpar,\xper)$ with
$\xpar=\lambda_0$ the coordinate of $x$ along $\LLc$ and $\xper\propto
u$ the coordinate along the chord $\xi(x_0)$. We choose $\xper$ so
that $d\xpar\wedge d\xper=dp\wedge dq$, which makes the change of
variables from $(p,q)$ to $(\xpar,\xper)$ canonical.

 In such a coordinate system, $\LLc$ is characterized by the equation $\xper=0$.  Moreover, if one considers a chord $\xi$ joining $\LL_+$
and $\LL_-$ and centered at a point of $\LLc$, $\xi$ is by definition
oriented in the transverse direction, implying that its parallel component vanishes, 
namely $\xipar=0$.
The endpoints $x^+$ and $x^-$ of the chord $\xi(x_0)$ going through a
point $x_0=(\xpar,0)$ are defined
by $\frac12(x^++x^-)=x_0=(\xpar,0)$ and $\xi=x^+-x^-$, and therefore we have
\begin{equation}
\label{coorx0}
x^{\pm}=(\xpar,\pm\frac12 w(\xpar))
\end{equation}
with $w(\xpar)=|\xi|$. Therefore, {\em in the units specified by the
  measure $d\xper$},  the function $w(\xpar)$ gives the width of the
region of small chord lengths at point $x_0$.

\subsubsection{Semiclassical integration in the transverse direction}
\label{sec:transverse}

We are now in a position to perform the integral \eqref{integralesimple} in the transverse direction,
for a fixed value of $\xpar$. As $\partial \Sx/\partial x = J\xi$, we have
$\partial \Sx/\partial \xper=\xipar$, so that the stationary condition
$\partial \Sx/\partial \xper=0$ is given by $\xipar=0$, that is,
stationary points lie on $\LLc$.

Fixing a point $x_0=(\xpar,0)$ on $\LLc$, we can perform the integral along $\xper$ in the
stationary phase approximation. By construction the stationary point is $\xper=0$,
and Gaussian integration yields
\begin{equation}
\int_{-\infty}^\infty d\xper e^{\frac{i}{2\hbar}\left(\frac{\partial
      \xi_\parallel}{\partial \xper}\right)_{\xper=0}
  \xper^2}=\sqrt{\frac{2i\pi\hbar}{\left(\partial
      \xi_\parallel/\partial \xper\right)_{\xper=0} }}\,. 
\end{equation}
Let us now compute the derivative $(\partial \xi_\parallel/\partial \xper)_{\xper=0}$. At an arbitrary point $x$ in the neighborhood of $x_0$, the endpoints $x^+(x)$ and $x^-(x)$ of the chord $\xi(x)$ going through $x$ coincide with endpoints of chords going through points lying on $\LLc$. More precisely, there exists some $\lambda_+ \in \LLc$ such that $x^+(x)=x^+(\lambda_+,0)$, and some $\lambda_-\in \LLc$ such that $x^-(x)=x^-(\lambda_-,0)$. As a consequence,
Eq.~\eqref{coorx0} yields
\begin{equation}
\label{xlambda}
x=\frac{x^+(x)+x^-(x)}{2}=\frac{x^+(\lambda_+,0)+x^-(\lambda_-,0)}{2}
=\left(\frac{\lambda_++\lambda_-}{2},\frac{w(\lambda_+)-w(\lambda_-)}{4}\right),
\end{equation}
and the corresponding chord $\xi$ reads
\begin{equation}
\label{xilambda}
\xi(x)=x^+(x)-x^-(x)=\left(\lambda_+-\lambda_-,\frac{w(\lambda_+)+w(\lambda_-)}{2}\right).
\end{equation}
Since both $\lambda_+$ and $\lambda_-$ are close to $x_0$ we can 
expand to first order 
\begin{equation}
\xper=\frac{w(\lambda_+)-w(\lambda_-)}{4}\simeq\frac{\lambda_+-\lambda_-}{4}w'(\xpar)=\frac{\xipar}{4}w'(\xpar),
\end{equation}
which yields
\begin{equation}
\left(\frac{\partial \xipar}{\partial \xper}\right)_{\xper=0}=\frac{4}{w'(\xpar)}\,.
\end{equation}
We thus obtain for the result of the transverse integration at fixed $\xpar$
\begin{equation}
\label{apresstatph}
\int d \xper \cA(x) e^{\frac{i}{\hbar}\Sx(x)} =
\frac12\cA(\xpar,0)e^{\frac{i}{\hbar}\Sx(\xpar,0)}\sqrt{2\pi\hbar |w'(\xpar)|}e^{i\frac{\pi}{4}\textrm{sgn}(w'(\xpar))} \; ,
\end{equation}
with $\cA$ defined by Eq.~\eqref{eq:def-cA}.

\subsubsection{Integration in the parallel direction}

To proceed with the last integral, one should bear in mind that the
parameterization of the Lagrangian manifold $\LL_t$ is essentially
arbitrary, and that any parametrization $\bs'$ of $\LL_t$ can be used
as long as $a(\bs)$ transforms into $ a(\bs')$ according to \eqref{eq:stosp}.

Therefore, instead of using
parameters $s^+$ and $s^-$ in the regions around $x^+$ and $x^-$ we
can use $\xpar$ as a parameter for both, with $\xpar$ the coordinate
on $\LL_c$ such that the endpoints of the chord going through
$(\xpar,0)$ coincide with $x^+$ and $x^-$ (note that $\xpar$ coincides
with $\lambda_+$ and $\lambda_-$ introduced above for $s^+$ and $s^-$
respectively). We shall denote by $a_+$ and $a_-$ the amplitudes in
the regions around $x^+$ and $x^-$, so that
$a_+(\xpar)\equiv a(s^+) \sqrt{|ds^+/d\xpar|}$ and
$a_-(\xpar) \equiv a(s^-) \sqrt{|ds^+/d\xpar|}$. With that
parametrization the prefactor in Eq.~\eqref{eq:SC-Wigner} is
proportional to
$\overline{a_-(\xpar)}a_+(\xpar)/\sqrt{\left|D_\parallel\right|}$ with
$D_\parallel$ expressed by \eqref{matrixD} but in the new variables,
that is,
\begin{equation}
\label{matrixDxpar}
D_\parallel
=\left(\frac{\partial
    \xpar^+}{\partial \xpar}\right)^T   \left(\frac{\partial
\xper^-}{\partial \xpar}\right) - 
\left(\frac{\partial
    \xper^+}{\partial \xpar}\right)^T   \left(\frac{\partial
\xpar^-}{\partial \xpar}\right) \, .
\end{equation}
Here, $x^\pm=(\xpar^\pm,\xper^\pm)$ is given by Eq.~\eqref{coorx0}, so that
${\partial x^\pm}/{\partial \xpar} = (1,\pm \frac12 w'(\xpar))$, which inserted in \eqref{matrixDxpar}
yields $D_\parallel=w'(\xpar)$. 
Introducing the notation $O(\xpar) \equiv O(x(\xpar,0))$ and
$\cA(\xpar) \equiv \cA(x(\xpar,0))$ (and for later use 
$\Sx(\xpar)\equiv \Sx(x(\xpar,0))$) we thus have
\begin{equation} \label{eq:cA-1d}
\cA(\xpar) = \sqrt{\frac{2}{\pi\hbar}} \, \frac{\overline{a_-(\xpar)}a_+(\xpar) 
e^{i \eta \pi/4-i\mu\pi/2}}{\sqrt{\left| w'(\xpar) \right|}}O(\xpar) \; .
\end{equation}
The factor $1/\sqrt{\left|  w'(\xpar) \right|}$ in \eqref{eq:cA-1d} thus exactly cancels the $\sqrt{\left| w'(\xpar) \right|}$ in \eqref{apresstatph} originating from the transverse integral. Taking into account the summation over all possible chords labeled by $\beta$ we finally get  
\begin{equation}
\label{eq:final}
\langle \hatO \rangle = \langle \hatO \rangle_{\rm classic}
+\sum_{\beta = \text{short chord}} \int d\xpar\, O({\xpar})
\overline{a_-^\beta({ \xpar})}a_+^\beta({ \xpar}) e^{\frac{i}{\hbar} \Sx_\beta({\xpar}) - i\mu_\beta \frac{\pi}{2}+i \eta_\beta \frac{\pi}{4}+i\eta^\perp_\beta\frac{\pi}{4}} +\textrm{ c.c.} \; ,
\end{equation}
where $\mu_\beta$ is the Maslov index of any path joining $\LL_+^\beta$ to
$\LL_-^\beta$ and $\eta^\perp$ is the sign of $w'(\xpar)$. 

Similarly to  the classical contribution Eq.~\eqref{eq:expectation}, the
remaining integration over $\xpar$ is actually very much
classical-like.  Indeed, because only phases that vary extremely
slowly contribute, all factors in the integrand vary only on the
classical scale, including the exponential term which contains an
$\hbar$ in its exponent.  This is therefore a rather atypical integral
in the context of semiclassical physics : it cannot be evaluated using
the separation of scales typical of the stationary phase approximation
(where anything varying only on the classical scale can be assumed
frozen in the relevant range of integration), which actually is the
reason why a blind application of stationary phase approximation fails
for the calculation of the expectation value of operators.

\subsection{Generalization to dimension  \texorpdfstring{$K$}{K}}

If we look back at the physical meaning of the calculation leading to
Eq.~\eqref{eq:final}, it can be summarized as follow.  First, we have
identified that the origin of the intereference terms corresponds to
pairs of sheets of the Lagrangian manifold that, under the classical
time evolution, become close to each other {\em on the quantum scale}.
On the classical scale they can therefore be considered as essentially
identical as $\hbar \to 0$, and one can therefore parametrize both of
them by the same coordinate $\bxpar$. Introducing then a transverse
direction $\bxper$ such that $(\bxpar,\bxper)$ forms a canonical pair
of variable, the integral along the transverse direction can be done
using stationary phase approximation, while the one on the parallel
direction does not involve any rapidly oscillating phase and is thus
essentially of a classical nature.

What eventually makes this calculation tractable is that we can apply
a canonical change of variable at various steps of the derivation and
place ourselves in the coordinate system for which the computation is
the simplest.  This is possible because, as dicussed in
section~\ref{sec:canonical}, the relevant contribution to the problem
we are considering are invariant under canonical changes of variable.

Having done this carefully for the one-dimensional case, we shall see now that the result \eqref{eq:final} can be obtained more directly and extended to arbitrary dimension if we start from the very definition \eqref{eq:Wignerpsi} of the Wigner function, with $\Psi_t$ the semiclassical wavefunction constructed from the propagated Lagrangian manifold $\LL_t$, and choose the right
coordinate system.  Again, as follows from the discussion of
Section~\ref{sec:canonical},  the relevant contributions to
\eqref{hata} of \eqref{eq:Wignerpsi}  can, too, be considered as canonical
invariants, which allows us to choose freely the coordinate system.

Consider thus a contribution to the interference terms associated with
a pair of sheets $(\LL^+,\LL^-)$ of $\LL$ which come extremely close
to each other.  On the classical scale these two sheets can be viewed
as identical, and because they are Lagrangian it is always possible to
find a system of canonical coordinates $(\bxpar,\bxper)$ such that
$\bxpar$ parametrizes $\LL^\pm$ (in the sense that
$(\bxpar,0) \in \LL^\pm$ if we do not distinguish between $\LL^+$ and $\LL^-$) and $\bxper$ is transverse to
it. Here however, and contrarily to the approach we used in
  section~\ref{sec:one-d} for the one-dimensional case, we use a
  coordinate system $(\bxpar,\bxsheer)$ such that $\bxsheer$ can,
  too, be used as a parametrization of $\LL$, so that it can
    play the role of $\bq$ in Eq.~\eqref{eq:Wignerpsi}.   A
  simple choice of coordinates  for such a requirement is the sheer
  $\bxsheer \equiv \bxper + \mu \bxpar$ (with $\mu \neq 0$
  arbitrary). We use from now on the variables $(\bxpar,\bxsheer)$.

In this coordinate system, the integral \eqref{hata} reads
\begin{equation}
\label{eq:Wignerpsistart}
\frac{1}{(2\pi\hbar)^K}\iiint \dK{\bxpar} \dK{\bxsheer} \dK{\bxsheer'} 
\cala(\bxsheer+\frac{\bxsheer'}{2})\overline{\cala(\bxsheer-\frac{\bxsheer'}{2})}O(\bxpar,\bxsheer)e^{-\frac{i}{\hbar}\bxpar.\bxsheer'+\frac{i}{\hbar} \Sq(\bxsheer+\frac{\bxsheer'}{2})- \frac{i}{\hbar} \Sq(\bxsheer-\frac{\bxsheer'}{2})}\,,
\end{equation}
where $O(\bxpar,\bxsheer)\equiv O(\bbx(\bxpar,\bxsheer))$. Changing
variables in the integral \eqref{eq:Wignerpsistart} from
$\bxsheer,\bxsheer'$ to $\bxsheer^+$, $\bxsheer^-$ with
$\bxsheer^\pm=\bxsheer\pm \frac12\bxsheer'$ (with Jacobian equal to
1), Eq.~\eqref{eq:Wignerpsistart} gives
\begin{equation}
\label{eq:Wignerpsistart2}
\frac{1}{(2\pi\hbar)^K}\iiint \dK{\bxpar} \dK{\bxsheer^+} \dK{\bxsheer^-} 
\cala(\bxsheer^+)\overline{\cala(\bxsheer^-)}O\left(\bxpar,\frac{\bxsheer^++\bxsheer^-}{2}\right)e^{-\frac{i}{\hbar}\bxpar.(\bxsheer^+-\bxsheer^-)+\frac{i}{\hbar} \Sq(\bxsheer^+)- \frac{i}{\hbar} \Sq(\bxsheer^-)}.
\end{equation}
We then use the fact that $\LL$ can be parametrized locally by an
arbitrary parameter $\bs$ and that the amplitude $\cala(\bxsheer)$ of
the semiclassical wavefunction is related to the density $a(\bs)$ on
$\LL$ by
$\cala(\bxsheer)=a(\bs)/\sqrt{|\partial\bxsheer/\partial\bs|}$. Choosing
$\bs=\bxpar$ as a local parameter in the vicinity of points $\bbx^+$
and $\bbx^-$ we have
\begin{equation}
\label{calap}
\cala(\bxsheer^+)=\frac{a(\bxpar^+)}{\sqrt{|\partial \bxsheer^+/\partial\bxpar^+|}},\qquad
\cala(\bxsheer^-)=\frac{a(\bxpar^-)}{\sqrt{|\partial \bxsheer^-/\partial\bxpar^-|}}\,.
\end{equation}
We now perform the stationary phase integration over the two
  variables $\bxsheer^+$ and $\bxsheer^-$ in
  \eqref{eq:Wignerpsistart2}. The critical points
  $(\bxsheer^{+*},\bxsheer^{-*})$ are those where the gradient of the
  phase vanishes, yielding the two conditions $\bxpar^+ (\bxsheer^{+*})
  = \bxpar$ and $\bxpar^-(\bxsheer^{-*}) = \bxpar$, corresponding
  respectively to the points on $\LL^+$ and $\LL^-$ parameterized by
  $\bxpar$.  The Hessian matrix is a block-diagonal matrix with blocks
  given by $(\partial \bxpar^+/\partial \bxsheer^+)$ and $(\partial
  \bxpar^-/\partial \bxsheer^-)$. Gaussian integration thus gives an amplitude
\begin{equation}
\label{eqzz}
(2\pi\hbar)^K\frac{1}{\sqrt{\Big|\left(\frac{\partial \bxpar^+}{\partial\bxsheer^+}\right)\left(\frac{\partial \bxpar^-}{\partial\bxsheer^-}\right)\Big|}}\,.
 \end{equation}
Inserting \eqref{calap}
and \eqref{eqzz} into Eq.~\eqref{eq:Wignerpsistart2} at the stationary point $\bxpar^+=\bxpar^-=\bxpar$, we get 
\begin{equation}
\label{eq:finalK}
\langle \hatO \rangle = \langle \hatO \rangle_{\rm classic} +
\sum_{\beta = \text{short chord}} \int \dK{\bxpar} \, O(\bxpar)
\overline{a_-^\beta(\bxpar)}a_+^\beta(\bxpar)
e^{\frac{i}{\hbar} \Sx_\beta(\bxpar)-i\theta_\beta\frac{\pi}{4}}+\textrm{ c.c.}  \: ,
\end{equation}
where (as in Sect. \ref{wignerstatphase}) each chord corresponds to two stationary points for which the roles of $\bxsheer^{+*}$ and $\bxsheer^{-*}$ are exchanged, yielding two complex conjugated terms.
In this equation,
\begin{equation}
  O_\beta(\bxpar) \equiv
  O_\beta\left(\bxpar,\frac{\bxsheer^+(\bxpar)+ \bxsheer^-(\bxpar)}{2} \right) \; ,
 \end{equation}
is the   Wigner transform of the operator $\hat O$ evaluated in $\bxpar$ at
the mid-point between $\LL^+$ and $\LL^-$;
\begin{equation}
\Sx_\beta(\bxpar) \equiv 
\Sx_\beta \left( \bxpar,\frac{\bxsheer^+(\bxpar)+ \bxsheer^-(\bxpar)}{2} \right)
 \end{equation}
  is the action Eq.~\eqref{Sxbx} evaluated at the same location, i.e.~the action integral on the contour formed by a path on $\LL$ from
  $\bbx^+_\beta$ to $\bbx^-_\beta$ which is closed by a straight line
  between $\bbx^-_\beta$ and $\bbx^+_\beta$; and $\theta_\beta$ is a
  global index. With this parameterization, the fact that the
  remaining integral on $\bxpar$ cannot be performed within a
  stationary phase approximation approach becomes obvious.  For $K=1$
  this expression coincides with Eq.~\eqref{eq:final}.

\begin{figure}[!t]
\begin{center}
\includegraphics[width=0.45\linewidth]{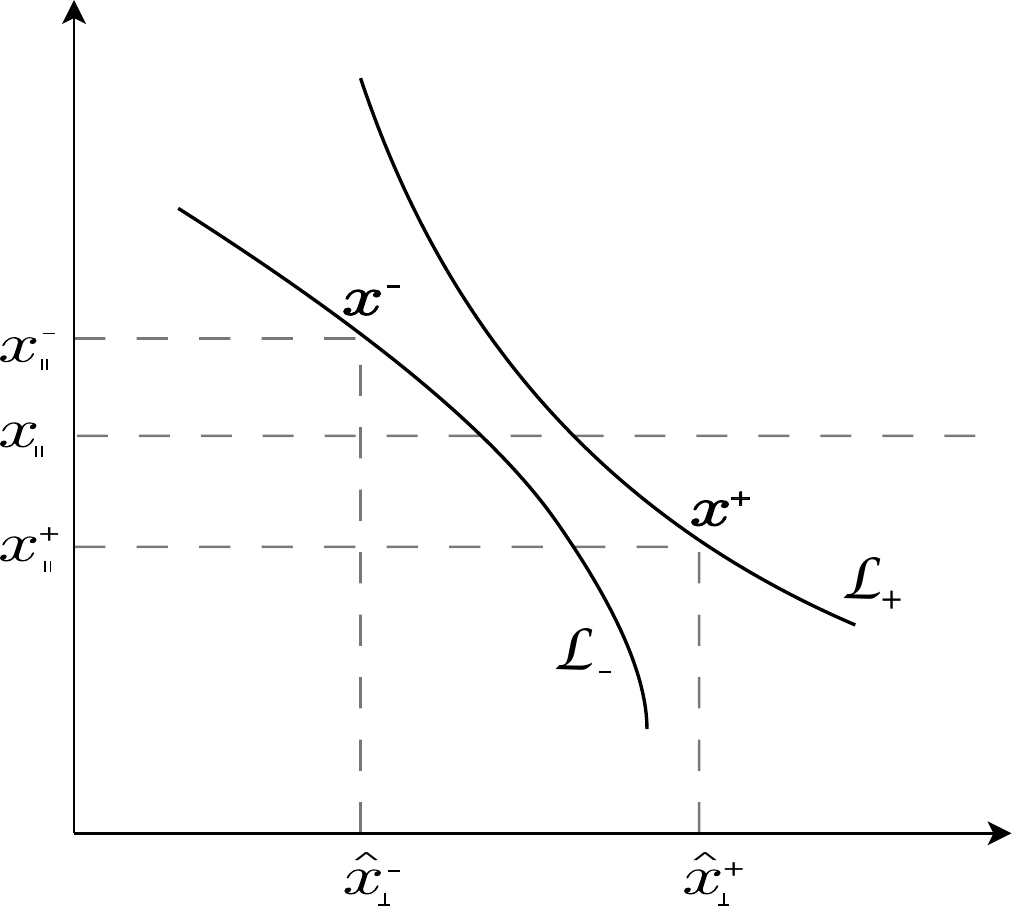}
\caption{
Parametrization used for the integral
\eqref{eq:Wignerpsistart}.  The integration variables are $(\bxsheer^+,\bxsheer^-,\bxpar)$, which all can be used (locally) as parametrization of  $\LL_\pm$. The stationary phase conditions reads  $\bxpar^+(\bxsheer^+) = \bxpar = \bxpar^-(\bxsheer^-)$.\label{figinteg}}
\end{center}
\end{figure}

The index $\theta_\beta$ in \eqref{eq:finalK} results from a contribution from the double Gaussian integration \eqref{eqzz} over $\bxsheer^+$, $\bxsheer^-$ performed at fixed $\bxpar$ (see Fig.~\ref{figinteg}), and from Maslov indices appearing in the definition of the semiclassical wavefunction constructed on the Lagrangian manifold $\LL_t$. The stationary phase condition in the Gaussian integration is met when points $\bbx^+$ and $\bbx^-$ are such that $\bxpar^+=\bxpar^-=\bxpar$. At that point, since $\LL_+$ and $\LL_-$ are almost parallel, the two blocks $(\partial \bxpar^+/\partial \bxsheer^+)$ and $(\partial \bxpar^-/\partial \bxsheer^-)$ of the Hessian matrix behave in the same way, so that the phase arising from Gaussian integration is zero. The total phase contribution in \eqref{eq:finalK} thus reduces to the  Maslov indices of the wavefunctions at the stationary phase points, i.e.~$\theta_\beta = \mu_\beta$.

\subsection{Discussion}
\label{sec:discussion}
In general, the specific characteristics of the initial density
$a(\bs)$, the shape of the evolved manifold $\LL_t$, and the specific
form of the Wigner transform $O(\bbx)$ of the considered operator, all
affect the final result, and it is not possible to provide an
expression for the last integral on $\bxpar$ that would be valid on a
completely general basis. In many circumstances, one would have to
resort to numerical integration to perform this last step. On the
other hand, and for the same reason stationary phase cannot be
applied, this is a ``simple'' integral which, when computed
numerically, does not require a particularly fine grid to achieve good
precision.

Going back to $K=1$, further progress can be made however in
two limiting cases that we discuss further now.  The first one
corresponds to times where the first interference contributions start
to appear (i.e.\ $t \simeq t_{\rm s.c.}$).  In that case, short
chords are not yet a typical feature of the evolved Lagrangian
manifold, and will generically correspond to a bottleneck near a point
$\bx0$ where the two sheets are closest.  In that case one recovers
the rather typical situation where it is the variation of the phase
$\Sx(\xpar)$ which cuts off the integral on $\xpar$.  Using the fact
that $\partial \Sx/\partial \xpar=\xi_\perp$, and expanding up to order
3 we get
\begin{equation}
\Sx(\xpar)=\Sx(x_0)+w_0 (\xpar -  x_0) +\frac{w''_0}{6}(\xpar -x_0)^3 \,,
\end{equation}
since at the bottleneck point $\bbx_0$ the length $w$ reaches a minimum and thus $w'_0=0$.
If we take the convention that $w'(0)>0$ for $\xpar>0$ we then have
\begin{equation}
\int_{-\infty}^\infty d\xpar
e^{\frac{i}{\hbar}(w(0)\xpar+\frac{w''(0)}{6}\xpar^3)}=  \left(
\frac{2\hbar}{w''_0} \right)^{1/3} \Ai \left[ w_0
  \left(\frac{2\hbar}{w''_0} \right)^{1/3} \right] \; ,
\end{equation}
and the integral in \eqref{eq:final} reduces to 
\begin{equation}
\langle \hatO \rangle_\beta =  a_+^\beta(x_0) a_-^\beta(x_0)
O(x_0)e^{ - i\mu_\beta \frac{\pi}{2}+i \eta_\beta \frac{\pi}{4}+i\eta^\perp_\beta\frac{\pi}{4}}   \left(
\frac{2\hbar}{{w''}_0^\beta} \right)^{1/3} \Ai \left[ w^\beta_0
  \left(\frac{2\hbar}{{w''}_0^\beta} \right)^{1/3} \right]
\; .
\end{equation}

The second limiting case we shall consider will on the other hand
correspond to the long times limit $t \gg t_{\rm s.c.}$ of a chaotic
dynamics, for which we further assume that the region ${\rm supp}(O)$ inside which
$O(\bbx)$ is significant, although large on the quantum scale,
remains relatively small on the classical one, and in particular on
the scale on which the stable and unstable manifolds of the classical
motion change significantly their shape.
In that case, because $t \gg t_{\rm s.c.}$, we can assume that the
various pieces of the evolved Lagrangian manifold $\LL_t$, and in
particular $\LL_+$ and $\LL_-$, essentially align on the unstable
manifold of the dynamics, and therefore remain essentially parallel
when they go through ${\rm supp}(O)$.  Furthermore, because of the
exponential stretching along the unstable manifold, the points of
$\LL_\pm \cap {\rm supp}(O)$ all originate from a very narrow neighborhood
of the original manifold $\LL_0$, and we can assume $a_\pm(\xpar) =
a_\pm = {\rm const.}$.  In that case, Eq.~\eqref{eq:final} reduces to 
\begin{equation}
\langle \hatO \rangle_\beta = \overline{a_-^\beta} a_+^\beta e^{\frac{i}{\hbar}\Sx_\beta(x_0) - \mu \frac{\pi}{2}}
\tilde O(w_\beta/\hbar)
\end{equation}
 (with $w_\beta$ the (constant) spacing between $\LL_+^\beta$
and $\LL_-^\beta$), and is thus expressed in terms of the Fourier component
\begin{equation}
\tilde O(w_\beta/\hbar) = \int d \xpar O(\xpar)e^{\frac{i}{\hbar} w_\beta (\xpar- x_0)}
\end{equation}
of $O(\xpar)$, i.e.\ of a cut of $O(x)$ along an unstable manifold.
Note that a rescaling of $\xpar$ would not change that result. \rem{t[o be
  checked though.]}

\section{Conclusion}
\label{ccl}

We have obtained a semiclassical expression for the mean value of
operators as a sum of a classical contribution
\begin{equation} \label{eq:TWAccl}
\langle \hat O \rangle_{\rm cl}(t) = \int d \bbx\, W_0(\bbx)\, O(\gt\bbx)
\end{equation}
corresponding to the TWA/LSC-IVR, and oscillatory terms associated with {\em
  short chords}. The final expression reads
\begin{equation}
\label{eq:finalKccl}
\langle \hatO \rangle = \langle \hat O \rangle_{\rm cl}(t)
+\sum_{\beta} \int d\bxpar\, O(\bxpar)
\overline{a_-^\beta(\bxpar)}a_+^\beta(\bxpar)
e^{\frac{i}{\hbar}\Sx_\beta(\bxpar)-i \mu_\beta \frac{\pi}{4}} +\textrm{ c.c.} \; ,
\end{equation}
with $\Sx_\beta(\bxpar)$ the action \eqref{Sxbx} evaluated at the mid-point between $\bbx^+(\bxpar)$ and $\bbx^-(\bxpar)$ (the points
with parameter $\bxpar$ on the sheets $\LL^+$ and $\LL^-$ of $\LL$),
$O(\bxpar)$ the value of the Wigner transform of the operator $\hat O$
at the same location, and $\mu_\beta$ the  Maslov index of the path joining  $\LL^+$ and $\LL^-$ on which $\Sx$ is computed. Here, we have used
the two invariances of the semiclassical Wigner function, invariance
under canonical change of variables (again in the
  approximation where only short chords contribute) and invariance
under parametrization of $\LL$, in order to choose $\bxpar$ both as
a  local coordinate and as a variable
parametrizing $\LL$.  Although this is
  not made explicit in our notations, different parametrizations
  $\bxpar$ can be used for different contributions $\beta$.

Our results show that time evolution of mean values of operators in
the semiclassical approximation cannot be obtained merely from the
stationary phase approximation, at least when several ingredients are
present. If the Wigner transform of the operator varies only at a
classical scale, time evolution may stretch the manifold on which the
semiclassical wave function is constructed in such a way
that non-stationary points become relevant to the semiclassical
dynamics. In that case, the stationary points (and their vicinity)
only provide the classical contribution to the mean value. These
stationary points correspond to cases where the chords joining a pair
of points on the Lagrangian manifold become of length zero. The semiclassical contributions originate in the fact that small (at the quantum scale) but nonzero chords may arise, for instance when the dynamics is chaotic. These almost stationary points do not lie in the vicinity of a truly stationary point, so that the stationary phase approximation does not capture them.

Thus, when computing semiclassically the mean values of smooth operators for a semiclassical wavefunction one needs to bear in mind that, when {\em short chords} are involved, i.e.\ when the folding of the Lagrangian manifold $\LL$ on which the wavefunction is built
induces the presence of sheets that are extremely close from one
another, $K$ integrals (over $\bxsheer$ in our notation) can be
performed within the stationary phase approximation; however the $K$
remaining ones (over $\bxpar$) do not involve fast oscillations or
stationary phase points and must be performed by other means. The
approach we suggest is thus to perform within the stationary phase
approximation all the integrals that can be done in this way, and to
perform the last remaining $K$ by other means, possibly numerically.
One implication of this approach however is that it makes
is  necessary to identify on the initial manifold (at $t=0$) which
  points will end up close to each other at time $t$, which amount to
  a root search for the classical dynamics.  This has to be contrasted
  with phase-space Initial (or Final) Value Representation (IVR or
  FVR)  approaches which are specifically designed to
  avoid such root search.

  These IVR/FVR semiclassical approaches are extremely popular in the
  chemical and molecular physics community, in particular in the
  version introduced by Herman and Kluck \cite{HermanKluk84} which
  involve some degree of smoothing, but also in the original form
  introduced by Miller \cite{Miller70} for the van Vleck propagator or
  in the more modern and elegant form for the Wigner function
  introduced by Ozorio de Almeida and coworkers \cite{Ozo13}.
  In these approaches, the classical trajectories implied in the
  semiclassical calculation under consideration are entirely specified
  by their initial conditions, thus avoiding the need of a root
  search.  As a bonus one moreover usually avoids the divergences associated with
  caustics or turning points.  The section 6 of \cite{Ozo13} gives for
  instance an application to these IVR/FVR approaches to the
  calculations of mean values of operators that we have analyzed in
  this paper.

If the initial state $\ket{\Psi_0}$ that one is propagating does not
correspond to a {\em  semiclassical wavefunction} (in the form we
define them in section~\ref{sec:SemiclassicWF}), there is no ambiguity
that the IVR/FVR form of the semiclassical propagation of the mean
value is superior to any approach that would imply a root search.  If
on the other hand $\ket{\Psi_0}$ does take the form of a semiclassical
state, then  IVR/FVR expressions 
    for the operator mean value are of course equivalent (within stationary
    phase approximation) to inserting the semiclassical expression
    \eqref{eq:SC-Wigner} into \eqref{hata} \cite{MGU-2020b}.  As
    expected, both describe  the same physics, and
    deciding on one or the other is mainly a matter of effectiveness;
    the term of the choice being  to do a root search or to avoid
    this root search at the cost of performing numerically extra
    integrals. 

  There is obviously a set of situations where avoiding the root
  search is enough of a simplification to justify the extra numerical
  integrals implied by the IVR/FVR approaches.  However, if we go back
  to the calculations of transverse integral in
  section~\ref{sec:transverse} in the simple one-dimensional case, we
  see that this integral is dominated by a region of size
  $O(\sqrt{\hbar w'})$, with $w'$ the variation of the distance
  between the two sheets of the manifold. As expected for
  semiclassical integrals evaluated at the stationary phase
  approximation, this size scales as $\sqrt{\hbar}$, and, as soon as
  one enters a bit deeply in the semiclassical regime, any attempt to
  compute numerically the integral will require a very fine grid to
  avoid the output being dominated by numerical noise.  This state of
  things is quite general in semiclassical integrals which in the
  semiclassical limit are usually dominated by very small portion of
  configuration (or phase) space.  Here, this is actually made worse
  by the factor $w'$, which tends to zero when the two sheets of the
  manifold become parallel, which will necessarily happen when they
  become very close to each other.  Even in the simple one-dimensional
  case, computing numerically the integral \eqref{integralesimple} in
  the deep semiclassical regime and for rather long evolution time of
  a chaotic system such that the folding of the manifold creates
  extremely close, almost parallel, pairs of sheets, will require to
  extract numerically the relevant information from a tiny area of
  phase space which, if the different role of the parallel and
  perpendicular directions are not recognized, will require to
  propagate classical trajectories on a grid on the sub-quantum scale
  in both directions.  On the other hand our approach would in that
  case only require a one-dimensional numerical integration (along the
  manifold) with a grid on the classical scale.

    We would thus argue that, if IVR/FVR are clearly the
    best options for semiclassical evolution of the mean value of an
    operator taken on a non-semiclassical wavefunction, and could be a
    viable route for short time / not too deeply semiclassic problems,
    the root search implied by our approach would still remain a
    considerably simpler and more accurate option in the deep quantum
    long-time chaotic regime. 

As a  final remark, in this paper we concentrated on the simplest form of
  folding of the classical manifold where, under chaotic time
  evolution, the Lagrangian manifold stretches out with a curvature on
  the classical scale. However much more involved classical structures
  can appear, such as the whorls and tendrils discussed in
  \cite{BerBal78}.  For us, it is an open question to decide whether
  the approach we propose can be adapted to these more complicated
  geometries, whether they could still be addressed semiclassically
  but would require numerical integration of the semiclassical
  expressions (bringing us back then to IVR/FVR), or whether
  they merely lead to diffractive effects that cannot be handled
  semiclassically.

\rem{ $W$ is a canonical invariant (cf Berry and Ozorio insisting only on $(p,q)\to (-q,p)$)}

\rem{Alternative approach with Green function (see forthcoming publi), citer DitVivSan06 etc}

\section*{Acknowledgments}
 We are deeply thankful to  Alfredo Ozorio de   Almeida for the role he played in motivating this work, as well as  in the many exchanges we had at later stages of the process.
  
  This work was supported by the ANR COCOA project, grant  ANR-17-CE30-0024-02 of the Agence Nationale de la Recherche.

\bibliographystyle{apsrev4-1}
\bibliography{references}

\end{document}